# Personalized Fashion Recommendation with Image Attributes and Aesthetics Assessment

Chongxian Chen[†]   Fan Mo[‡]   Xin Fan[‡]   Hayato Yamana[§]

[†] [‡] Graduate School of Fundamental Science and Engineering 3-4-1 Okubo, Shinjuku-ku, Tokyo, 169-8555 Japan

[§] Faculty of Science and Engineering, Waseda University 3-4-1 Okubo, Shinjuku-ku, Tokyo, 169-8555 Japan

E-mail:   [†] chenc@toki.waseda.jp,   [‡] bakubonn@toki.waseda.jp,   [‡] fan_xin@fuji.waseda.jp

[§] yamana@yama.info.waseda.ac.jp

**Abstract** Personalized fashion recommendation is a difficult task because 1) the decisions are highly correlated with users' aesthetic appetite which previous work frequently overlooks, and 2) many new items are constantly rolling out that cause strict cold-start problems in the popular identity (ID)-based recommendation methods. These new items are critical to recommend because of trend-driven consumerism. In this work, we aim to provide more accurate personalized fashion recommendations and solve the cold-start problem by converting available information, especially images into two attribute graphs focusing on optimized image utilization and noise-reducing user modeling. Compared with previous methods that separate image and text as two components, the proposed method combines image and text information to create a richer attributes graph. Capitalizing on the advancement of the large language and vision models, we experiment with extracting fine-grained attributes ever efficiently and as desired using two different prompts. Preliminary experiments on the IQON3000 dataset have shown that the proposed method achieves competitive accuracy compared with baselines.

## 1. INTRODUCTION

Recommender systems have become indispensable tools in tackling the contemporary challenge of information overload. Recommender systems play a crucial role in enhancing customers' online shopping experiences by offering personalized lists of potential items. While previous approaches have achieved considerable success using graph neural networks (GCN) for personalized recommendations [8,9], the prevalent ID-based approach in GCN encounters difficulties with the cold-start problem. Recent research has demonstrated success in overcoming this challenge by leveraging item attribute graphs [1]. However, personalized fashion recommendation presents unique challenges due to the critical role of image-driven influence on shopping decisions. Previous methods in fashion recommendation have predominantly relied on pre-trained image models for extracting embeddings to capture users' image preferences.

In this study, we propose an innovative approach to leverage images in personalized fashion recommendations. Capitalizing on the advancements in large language and vision models (LLVM) [14], our method processes images into two distinct parts of fine-grained attributes ever efficiently and reproducibly. We claim that there are two distinct influences in images that affect users' shopping decisions. Firstly, item-specific attributes that describe features of an item, and secondly, image aesthetic attributes that capture users' aesthetic preferences. An example is shown in Fig. 1 to demonstrate the difference between item-specific attributes and image aesthetic attributes. Alongside the fine-grained keywords from item descriptions, we construct the Items and Item Attributes Graph that integrates both images and text from the items. To model the users' preferences, we construct a Users, Image Aesthetic Attributes and Items Graph to integrate not only item information but explicitly model users' aesthetic preferences.

Unlike other shopping domains such as groceries and hardware, fashion shopping places a significant reliance on images for decision-making. Additionally, the constant influx of new fashion items each season, coupled with consumers' high interest in new trends, underscores the critical role of fashion recommender systems in recommending these new items. Our proposed method integrates both images and texts to construct the Items and Item Attributes Graph and the Users, Image Aesthetic Attributes and Items Graph, effectively addressing the cold start problem and offering optimized utilization of images compared to previous methods.

Our approach addresses two primary challenges in personalized fashion recommendation: 1) the ability to recommend new items, i.e., solving the cold-start problem, and 2) the optimized utilization of image data to capture users' personal aesthetic preferences. Instead of treating text and images as separate entities, our proposed method combines both sources to construct a comprehensive attributes graph, providing a unique solution to personalized fashion recommendations compared with previous independent components. The proposed method also better models users' preferences explicitly by connecting users with not just items, but also image aesthetic attributes by constructing the innovative Users, Image Aesthetic Attributes and Items Graph. Previous methods [4,5] model users' preferences by integrating item information, which we claim to have more item-specific noise that is not related to the users' preferences.

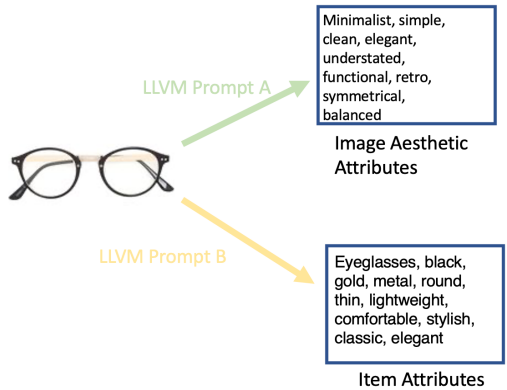

Fig. 1. A demonstration of example Image Aesthetic Attributes and Item Attributes.

The proposed methods contain three major components to model the items: 1) extract item attributes from items' images using LLVM. Utilizing the advancement in LLVM, we are able to experiment with extracting fine-grained attributes from the images more efficiently and as desired, and 2) extract item attributes from items' text descriptions. The brand, price, material, category, color, and introduction are carefully processed into item attributes, and 3) construct the Items and Item Attributes Graph. With item attributes extracted from both images and texts, we ensure to model the items comprehensively.

We also propose a novel method to model the users in order to reduce the items' noise in representing users' preferences. We utilize LLVM to extract image aesthetics from the images to model users' aesthetic preferences. Together with item embeddings learned from the Items and Item Attributes Graph, we build the Users, Image Aesthetic Attributes and Items Graph.

The main contributions of this paper are summarized as 1) we are the first to experiment with constructing an Items and Item Attributes graph using both image and text information, 2) we show the results of experimenting with image attributes extraction using two different prompts using upgraded LLVM, 3) an innovative approach to model the users' preference to reduce the item-specific noise.

## 2. RELATED WORK

### 1. Graph Neural Networks and the Cold Start Problem in the Recommender Systems

Graph Neural Networks (GNN) have shown success in recommender systems. Recent graph-based recommenders [2, 3] typically construct user-item bipartite graphs and then capture user-item correlations by propagating their embeddings using GNNs [8, 9]. The powerful information propagation using graphs enables efficient and large-scale modeling of the items and users. However, such methods suffer from the cold start problem that the dominating ID-based approach fails to work. The new items are missing the ID-based embeddings because the models are not trained on these new items, thus being unable to recommend the new items. To solve the cold start problem, Cao et al. [1, 13] proposed to construct an item-attributes graph. However, Cao et al.'s [1] fine-grained methods require careful preprocessing in extracting the attributes from text.

### 2. Fashion Recommender Systems

The compatibility between fashion items has attracted studies in recent years. Effectively modeling the compatibility between fashion items is challenging in fashion research [4,5,7]. Traditionally, this problem has been cast as a metric learning challenge, wherein a compatibility space is learned, and the distance between items within this space reflects their compatibility [4]. Recent studies [5] has been utilizing hierarchical graphs to model the relationship between users, sets, and items. Despite the effectiveness, recent studies [4,5,7] have been utilizing pre-trained image models like ResNet to extract visual features. Such approaches separate the text attributes from the image, making them two independent components in modeling fashion items, causing nonoptimal performance, also demonstrated by the ablation study. In our proposed methods, we aim to combine images and texts as one component for optimal utilization of all data sources to model fashion items.

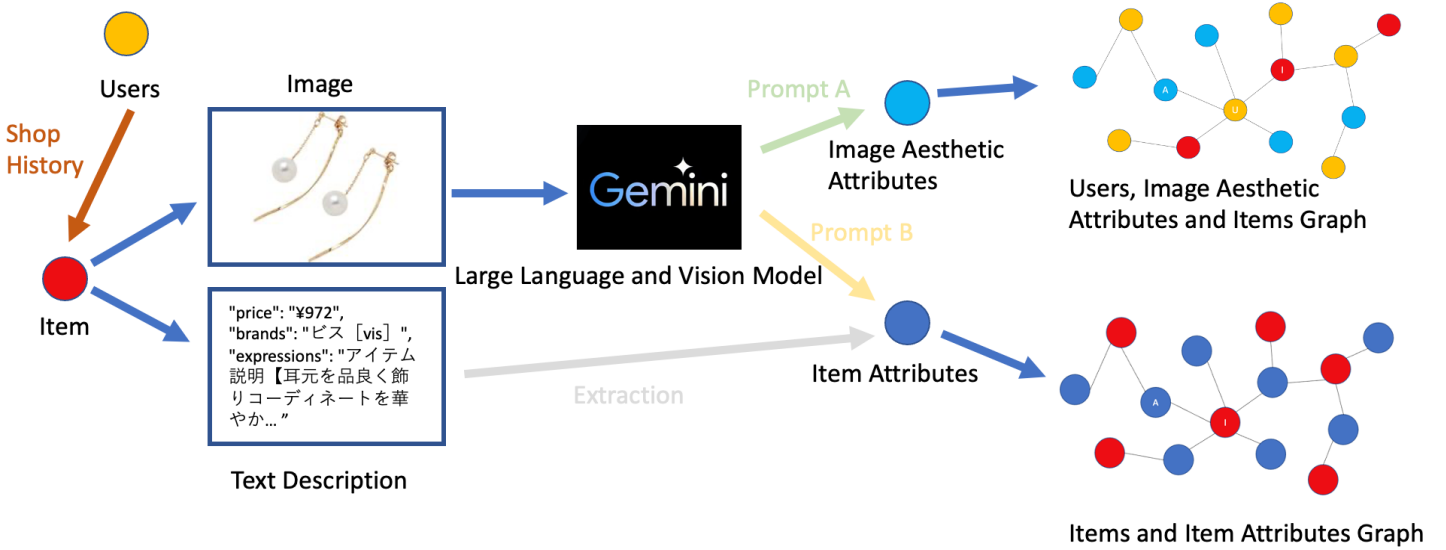
Fig. 2. An overview of the Items and Item Attributes Graph and the Users, Image Aesthetic Attributes and Items Graph construction.

*Gemini is a trademark of Google LLC.

## 3. PRELIMINARIES

Before presenting our proposed methodology, we provide an overview of GCN-based Collaborative Filtering.

### 1. GCN-based Collaborative Filtering

Graph Convolutional Networks (GCNs) [2,3] are designed to capture collaborative signals within historical interaction data through convolutional operations on the user-item bipartite graph constructed from past user-item interactions. In the forward propagation phase, given the initial representations $e_u$ and $e_i$ for user u and item i, GCNs aggregate information from neighbors recursively to create final representations, and $e_i$, for user u and item i, respectively. These representations are then utilized to calculate the preference score representing the user's preference for item i. Subsequently, the preference scores for unseen items are used to generate the recommendation list.

During the backpropagation process, the parameters of GCNs are optimized using the Bayesian personalized ranking (BPR) loss [10]. This loss function is pairwise, ensuring that a user's preference score for interacted items is higher than the preference score for non-interacted items. In conclusion, GCNs learn to enhance recommendation performance by iteratively updating their parameters based on the information propagation captured from graphs made from historical interactions, providing personalized, effective, and large-scale recommendations for users.

## 4. METHODOLOGY

In this section, we elaborate on the detailed proposed methods. Table I. shows the symbols used in this paper.

### A. Image to Item Attributes Extraction

Instead of treating the image as just embedding, we propose to adopt LLVM to extract attributes from the image ever efficiently and as desired to enrich our attributes graph. With the recent advancement in LLVM, we experiment with Google's Gemini Pro Vision model for image attributes extraction. Each time we provide an item's image and a prompt to the model using Google's Python API google.generativeai. After experimenting, the prompt we used to extract the item attributes from images is named Prompt B as follows: "Describe the item in the image using keywords. Describe the color, material, pattern, style, and feeling of the item using simple, common, and many English keywords. Output as keyword1, keyword2, keyword3, ... keywordn." The results are satisfying with an example demonstrated in Fig. 1.

TABLE I. SYMBOLS

| Symbol | Explanation |
|---|---|
| $e_i$ | Item embedding |
| $e_{ia}$ | Item Attributes embedding |
| $e_{iaa}$ | Image Aesthetic Attributes embedding |
| $e_u$ | Users embedding |
| $N_{i\_u}$ | the set of items that were interacted by user $u$ |
| $N_{iaa\_u}$ | the set of image aesthetic attributes that were interacted by user $u$ |
| $N_{ia\_i}$ | the set of item attributes that were interacted by item $i$ |
| $e_i^k$ | The embedding of items at the k-th layer |
| $e_u^k$ | The embedding of users at the k-th layer |
| $n_p$ | The number of total price categories in item attributes |
| $n_n$ | The number of negative item samples in the experiment |
| $G_{iia}$ | The Items and Item Attributes Graph where $G_{iia} = \{V_i, V_{ia}, E_{iia}\}$ |
| $G_{uiaai}$ | The Users, Image Aesthetic Attributes and Items Graph where $G_{uiaai} = \{V_u, V_i, V_{iaa}, E_{uiaai}\}$ |

### B. Text Description to Item Attributes Extraction

Each item includes a text description that includes the brand, price, category, color, and item description. The texts are carefully analyzed and processed to extract useful keywords as part of the item attributes. Because the prices are discrete values, we process the prices into $n_p$ categories, where $n_p$ is a tunable parameter.

### C. Image to Image Aesthetics Attributes Extraction

Besides extracting item attributes from the image, we claim that it is also important to extract image aesthetics to better model users' preferences. Users' shopping decisions are driven by the images they see. If we only aggregate the item embeddings to model the users, we will get unnecessary item-related noise. For example, a user may buy a pair of glasses with the "lightweight" attribute, but not necessarily prefer a "lightweight" sweater. We claim that users may have image aesthetics preferences like "symmetrical", thus it is important to extract image aesthetics attributes from the image. We used Prompt A: "Describe the image aesthetics independent of the item using keywords. Describe qualities including image composition, color scheme, lighting, balance, symmetry, contrast, texture, and overall visual harmony, feeling of the image using simple and common English keywords. Output as keyword1, keyword2, keyword3, ... keywordn" to extract the image aesthetics attributes in our experiment.

### D. Items and Item Attributes Graph

We construct the Items and Item Attributes bipartite graph where each item is connected with the associated attributes from the image and the text. We define the graph with vertices consisting of items ($V_i$) and attributes ($V_{ia}$), as $G_{iia} = \{V_i, V_{ia}, E_{iia}\}$, where $E_{iia}$ is the set of edges that connects $V_i$ and $V_{ia}$.

### E. Users, Image Aesthetic Attributes and Items Graph

We construct the Users, Image Aesthetic Attributes and Items Graph where each user is connected with the items from the users' history, and with the image aesthetic attributes associated with the items directly as shown in Fig. 2. We define the graph with vertices consisting of users ($V_u$), items ($V_i$) and image aesthetic attributes ($V_{iaa}$), as $G_{uiaai} = \{V_u, V_i, V_{iaa}, E_{uiaai}\}$, where $E_{uiaai}$ is the set of edges that connect $V_u$ and $V_i$, and the set of edges that connect $V_u$ and $V_{iaa}$.

### F. Graph Convolution Networks

We initialize $e_{ia}^0$, $e_{iaa}^0$, $e_i^0$, $e_u^0$ as trainable parameters to represent the embedding of item attributes, image aesthetic attributes, items and users respectively. The embedding of an item vertex in graph $G_{iia}$ at the $k+1$ layer is calculated by (1) to integrate neighbors' information.

$$e_i^{(k+1)} = \sum_{a \in N_{i\_ia}} \frac{1}{\sqrt{|N_{i\_ia}||N_{ia\_i}|}} e_{ia}^{(k)} \quad (1)$$

, where $N_{ia\_i}$ is the set of item attributes that is connected to item $i$ in graph $G_{iia}$ and $N_{i\_ia}$ is the set of items that is connected to attributes $ia$ in graph $G_{iia}$. Similarly, the embedding of an item attribute vertex in graph $G_{iia}$ at the $k+1$ layer is calculated by (2)

$$e_{ia}^{(k+1)} = \sum_{i \in N_{ia\_i}} \frac{1}{\sqrt{|N_{ia\_i}||N_{i\_ia}|}} e_i^{(k)} \quad (2)$$

The embedding of an image aesthetic attribute vertex in graph $G_{uiaai}$ at the $k+1$ layer is calculated by (3)

$$e_{iaa}^{(k+1)} = \sum_{u \in N_{iaa\_u}} \frac{1}{\sqrt{|N_{iaa\_u}||N_{u\_iaa}|}} e_u^{(k)} \quad (3)$$

| Data Set | #Users | #Items | #User-item interactions | #Train | #Validation | #Test | Image | Text |
|---|---|---|---|---|---|---|---|---|
| IQON3000 | 3,568 | 23,363 | 459,146 | 367,317 | 45,914 | 45,915 | One for each item | Brand, price, category, color, and a text introduction of 50-150 words. |

TABLE II. EXPERIMENTAL DATASETS

TABLE III.  EXPERIMENTAL EVALUATION RESULT

| Models | IQON3000 | | |
|---|---|---|---|
| | *Recall@50* | *NDCG@50* | *Precision@50* |
| SAERS [12] | 0.2202 | 0.1373 | 0.0178 |
| MM-FRec [11] | 0.3157 | 0.1732 | 0.0214 |
| Proposed Method | 0.2863 | 0.1577 | 0.0197 |

, where $N_{iaa\_u}$ is the set of image aesthetic attributes that is connected to user $u$ in graph $G_{uiaai}$ and $N_{u\_iaa}$ is the set of users that is connected to image aesthetic attributes $iaa$ in graph $G_{uiaai}$. The embedding of a user vertex in graph $G_{uiaai}$ at the $k+1$ layer is calculated by (4)

$$e_u^{(k+1)} = \sum_{iaa \in N_{u\_iaa}} \frac{1}{\sqrt{|N_{u\_iaa}||N_{iaa\_u}|}} e_{iaa}^{(k)} + \sum_{i \in N_{u\_i}} \frac{1}{\sqrt{|N_{u\_i}||N_{i\_u}|}} e_i^{(k)} \quad (4)$$

, where $N_{i\_u}$ is the set of items that is connected to user $u$ in graph $G_{uiaai}$ and $N_{u\_i}$ is the set of users that is connected to item $i$ in graph $G_{uiaai}$.

Then, the final embedding of a user $e_u$ is obtained by (5)

$$e_u = \sum_{k=0}^{K} \alpha_k e_u^{(k)} \quad (5)$$

, where $K$ is a tunable parameter. In this paper, we experimented with $K \in \{0,1,2,3,4,5\}$ according to He et al. [2]. Similiarly we obtain final embedding of an item $e_i$ by (6)

$$e_i = \sum_{k=0}^{K} \alpha_k e_i^{(k)} \quad (6)$$

### G. Model Training

This sub-section explains the training details of the proposed model. We use the items a user interacted with in the data set as the positive samples and randomly select $n_n$ items that the user does not interact with as the negative samples. The model parameters are trained to minimize the BPR loss of the model output compared with ground truth. L2 regularization is utilized to avoid model overfitting.

For each training user-item interaction, we forward $N_{i\_u}$, $N_{ia\_i}$, and $N_{iaa\_u}$ into the model $f_\theta$ as (7)

$$p = f_\theta(N_{i\_u}, N_{iaa\_u}, N_{ia\_i}) \quad (7)$$

where p is the predicted probability of the future interacted item. Then, the network parameters $\theta$ are trained to maximize the predicted probability $p$ compared with the ground-truth item.

## 5. EXPERIMENTAL EVALUATION

### A. Datasets

To evaluate the effectiveness of the proposed method, we conducted experiments on the publicly available real-world dataset: *IQON3000* [4]

- **IQON3000 [4]:** The IQON3000 dataset is collected from the fashion shopping website Iqon. The IQON3000 dataset comprises 308,747 outfits of 3,568 users, with a total of 672,335 items. To ensure a comparison with previous work by Song et al. [11], we filtered the dataset the same way, which is only items that are interacted by more than 10 users are left in the dataset. Because no actual train/test data are provided after splitting is provided by previous work, we randomly sample 80%, 10%, and 10% of the user-item interactions in the dataset to form the training set, validation set, and testing set. The statistics of the dataset after filtering are shown in Table II.

### B. Experimental Setup

We introduce two baselines below. Since the dataset is the same, and the train, sample, and test ratio is the same we adopt the reported experimental results by Song et al. [11] for comparison.

**SAERS [12]**: A multi-modal-based approach that characterizes the preferences of users from both the visual and text attributes perspectives.

**MM-FRec [11]**: A multi-modal-based approach that recommends fashion items using both visual and semantic attributes.

### C. Discussion

We show the experimental evaluation result in Table III. The experimental performance from the proposed methods is worse in the three metrics than the previous best-performing method MM-FRec [11]. We plan to study the attributes we extracted from the LLVM more carefully and experiment with other prompts to improve the performance of the proposed method. Also, we plan to do the ablation study to analyze the effectiveness of the components in the proposed methods to further refine our methods. Because the proposed method can solve the cold-start problem, we also plan to test the performance on the unfiltered datasets, which MM-FRec can not experiment on because of the cold-start problem.

## 6. CONCLUSION

In this work, we propose a novel way for fashion recommendation specifically aiming to solve the cold start problem and the insufficiency of utilization of images. By extracting two types of fine-grained attributes from the images using LLVMs and text processing, our method presents a method that is reproducible and efficient to use for large-scale recommendation compared with traditional fine-grained attribute extraction. Combining both the attributes from the image and text, the item attributes graph proposed by our method is richer, bringing more explainable recommendations. The innovative user modeling attempts to directly consider the aesthetic preferences of users and reduces the noise from item-specific features. For future work, we plan to do an ablation study to analyze the effectiveness of the components in the proposed method. Also, we have not considered the item compatibility between fashion items in this paper, which may be further utilized to improve the accuracy of fashion recommendations.

## ACKNOWLEDGEMENTS

This work was supported by JST SPRING Grant Number JPMJSP2128.